\documentclass[]{aa}  
\usepackage{graphicx}
\usepackage{natbib}
\bibpunct{(}{)}{;}{a}{}{,}
\usepackage{txfonts}
\newcounter{Rco}
\newcommand{\Ionst}[1]{\setcounter{Rco}{#1}\Roman{Rco}}
\newcommand{\Ion}[2]{\mbox{#1\ {\scriptsize\Ionst{#2}}}}
\newcommand{\Ionw}[3]{\mbox{#1\ {\scriptsize\Ionst{#2}}~$\lambda\,#3$\,\AA}}

\newcommand{\Ionww}[3]{\mbox{#1\ {\scriptsize\Ionst{#2}}~$\lambda\lambda\,#3$\,\AA}}

\newcommand{\logg}{\mbox{$\log g$}}
\newcommand{\loggw}[1]{\mbox{$\log g\hspace{-0.5mm} =\hspace{-0.5mm}  #1$}}

\newcommand{\ab}[1]{\mbox{Fig.\,\ref{#1}}}
\newcommand{\aba}[1]{\mbox{Figure\,\ref{#1}}}
\newcommand{\sA}[1]{\mbox{(Fig.\,\ref{#1})}}

\newcommand{\se}[1]{\mbox{Sect.\,\ref{#1}}}
\newcommand{\sK}[1]{\mbox{(Sect.\,\ref{#1})}}

\newcommand{\ta}[1]{\mbox{Table\,\ref{#1}}}
\newcommand{\sT}[1]{\mbox{(Table\,\ref{#1})}}
\newcommand{\Teff}{\mbox{$T_\mathrm{eff}$}}
\newcommand{\Teffw}[1]{\mbox{$\Teff\hspace{-0.5mm}=\hspace{-0.5mm}#1\,\mathrm{kK}$}}
\newcommand{\aador}{AA\,Dor}
\newcommand{\lb}{LB\,3459}
\def\tmap{\emph{TMAP}}
\def\ironic{\emph{IrOnIc}}
\def\owens{\emph{OWENS}}
\begin{document}
\title{FUSE spectroscopy of sdOB primary \\ of the post common-envelope binary \lb\ (\aador)\thanks
        {Based on observations made with the NASA-CNES-CSA Far Ultraviolet 
         Spectroscopic Explorer. FUSE is operated for NASA by the Johns Hopkins
         University under NASA contract NAS5-32985.
        }
\author{J\@. Fleig\inst{1}         \and 
        T\@. Rauch\inst{1}         \and 
        K\@. Werner\inst{1}        \and 
        J\@. W\@. Kruk\inst{2}
       }
}

\institute{Institute for Astronomy and Astrophysics,
           Kepler Center for Astro and Particle Physics,
           Eberhard Karls University, 
           Sand 1,
           72076 T\"ubingen, 
           Germany,
           \email{rauch@astro.uni-tuebingen.de}
           \and
           Department of Physics and Astronomy, Johns Hopkins University, Baltimore, MD 21218, U.S.A.
           }


\date{Received August 4, 2008; accepted September 5, 2008}


\abstract
         {\lb\ (\aador) is an eclipsing, close, post common-envelope binary (PCEB) consisting of an sdOB primary star  
          and an unseen secondary with an extraordinarily low mass ($M_2 \approx 0.066\,\mathrm{M_\odot}$) -- 
          formally a brown dwarf. 
          A recent NLTE spectral analysis shows a discrepancy with the surface gravity, which is derived from
          analyses of radial-velocity and lightcurves.
         }
         {We aim at precisely determining of the photospheric parameters of the primary, especially of the surface
          gravity, and searching for weak metal lines in the far UV.
         }
         {We performed a detailed spectral analysis of the far-UV spectrum of \lb\ obtained with
          FUSE by means of state-of-the-art NLTE model-atmosphere techniques.
         } 
         {A strong contamination of the far-UV spectrum of \lb\ by interstellar line absorption hampers a precise 
          determination of the photospheric properties of its primary star.
          Its effective temperature (\Teffw{42}) was confirmed by the evaluation of
          new ionization equilibria. 
          For the first time, phosphorus and sulfur have been identified in the spectrum of \lb.
          Their photospheric abundances are solar and 0.01\,times solar, respectively.
          From the \Ionww{C}{3}{1174-1177} multiplet, we can measure the rotational velocity 
          $v_\mathrm{rot}=35\pm 5\,\mathrm{km/sec}$ of the primary of \lb\ and confirm that the rotation is bound.
          From a re-analysis of optical and UV spectra (analogue to Rauch 2000), we determine a slightly
          higher surface gravity \loggw{5.3\pm 0.1} compared to Rauch (2000, \loggw{5.2\pm 0.1}).
         }
         {The rotational velocity of the primary of \lb\ is consistent with a bound rotation. 
          The higher \logg\ reduces the discrepancy in mass determination in comparison to
          analyses of radial-velocity and lightcurves. However, the problem is not completely solved.
         } 

\keywords{Stars: abundances -- 
          Stars: atmospheres -- 
          Stars: binaries: eclipsing --
          Stars: early-type --
          Stars: low-mass, brown dwarfs  -- 
          Stars: individual: \aador, \lb
}

\maketitle

\section{Introduction}
\label{sect:introduction}
The eclipsing binary system \lb\, (\aador) is a blue foreground object of the LMC at a spectroscopic distance of
\mbox{$d=396\,\mathrm{pc}$} \citep{Rauch00}. It consists of aa sdOB with an effective temperature of \Teffw{42} and a 
low-mass companion from which no direct spectroscopic information has been obtained yet. Due to its mass of about 
$0.066\,\mathrm{M_\odot}$ it formally lies in the brown-dwarf range. The orbital period is about 0.26\,d and the 
inclination is $i=90\degr$. 

The system was analyzed several times, starting with \citet{Kilkenny79,Kilkenny81} and \citet{Paczynski80}, 
who established the basic parameters. First NLTE analyses of the primary were performed by \citet{Kudritzki82} 
and \citet{Lynas-Gray84}. Details on the history of this object can be found in \citet{Rauch00, Rauch04}. 

Recent spectral analyses by means of NLTE model-atmospheres that were based on 
optical (ESO CASPEC\footnote{European Southern Observatory, Cassegrain Echelle Spectrograph}) and 
ultraviolet (IUE\footnote{International Ultraviolet Explorer}) observations 
\citep{Rauch00} have shown a discrepancy with analyses of radial-velocity and lightcurves
\citep{Hilditch96,Hilditch03}. \citet{Rauch00} determined a surface gravity of \loggw{5.21\pm 0.1}, while
\citet{Hilditch03} obtained \loggw{5.45 - 5.51} from lightcurve and mass function.
Because the analysis of \citet{Rauch00} suffered from the long exposure times (some hours) hence smearing, due to
the orbital movement of the available spectra,
\citet{RauchWerner03} measured the radial-velocity curve from optical spectra with short (180\,sec) exposure times
obtained with ESO's VLT\footnote{Very Large Telescope} and UVES\footnote{Ultraviolet and Visual Echelle Spectrograph}.
They obtained a radius $r_1=0.169\,\mathrm{R_\odot}$ which is smaller than the
$r_1=0.236\,\mathrm{R_\odot}$ found by \citet{Rauch00}. Since the stellar radius depends on $g$ 
this may be a hint of a higher value than \loggw{5.21}. However, data reduction of the Balmer lines in the 
UVES spectra used by \citet{RauchWerner03} was not very accurate, so a precise determination of \logg\ of the primary 
by means of NLTE modeling techniques is still lacking.  

Consequently, high-resolution and high-S/N observations (exposure times of 200\,sec each) in the far-UV range were performed 
with the FUSE\footnote{Far Ultraviolet Spectroscopic Explorer} satellite. The FUSE wavelength range 
($904\,\mathrm{\AA} < \lambda < 1187\,\mathrm{\AA}$) covers the
hydrogen Lyman series except for Ly\,$\alpha$. The series decrement is a sensitive indicator for \logg.

A detailed spectral analysis of the FUSE observation of \lb\ is described in \sK{sect:analysis}.
Since the far-UV spectrum of \lb\ turned out to be strongly contaminated by interstellar absorption,
we modeled the ISM line absorption \sK{sect:observations} in order to distinguish weak lines of 
iron-group elements (here: Ca, Sc, Ti, V, Cr, Mn, Fe, Co, Ni; \se{sect:analysis}).

\section{The far-UV spectrum of \lb}
\label{sect:observations}

The FUSE instrument consists of four independent, co-aligned telescopes and
spectrographs. Taken together, the four channels span the wavelength range
$904\,\mathrm{\AA} < \lambda < 1187\,\mathrm{\AA}$ with a typical resolving 
power of $R \approx 20\,000$.  Further information on the FUSE mission and
instrument can be found in \citet{moos2000} and \citet{sahnow2000}.

Far-UV observations were performed with FUSE on August 29, 2003 
(observation id: D0250101) and June 22, 2004
(id: D0250102) using the LWRS aperture with a resolving power of $R \approx 20\,000$.
The individual exposure times were about 200\,sec to minimize effects of orbital motion.
Problems with coalignment of the telescope channels in the second observation resulted in loss of
the SiC channel data and most of the LiF2 channel data; the total exposure times varied from
730\,sec in the SiC channels to 2\,335\,sec in LiF1. The data were reduced with CalFUSE v\@. 3.1.3, but
a subsequent reduction with the final version of CalFUSE, v\@. 3.2.2, did not result in any significant
changes to the spectrum.  A correction for the ``worm'' feature (a shadow cast by
the detector grid wires) in LiF1b was obtained from a highly-smoothed
ratio of the LiF2a to LiF1b spectra. For a representative discussion of
FUSE data reduction procedures, see \citet{kruk2002}, and \citet{dixon2007}
for a description of CalFUSE.

An additional far-UV observation was performed earlier with BEFS\footnote{Berkeley Extreme and Far-UV Spectrometer}
($R \approx 10\,000$) aboard ORFEUS\,II\footnote{Orbiting and Retrievable Far and Extreme Ultraviolet Spectrometer} on
November 30, 1996 (id: BEFS2162) with an exposure time of 1\,112\,sec. We retrieved it from 
MAST\footnote{Multimission Archive at the Space Telescope Science Institute}. In \ab{fig:ORFEUS} we show a comparison
of ORFEUS and FUSE observations. We note that the measured flux levels
agree very well. For our analysis we mainly used the FUSE observation that has better resolution
and S/N. All observations are slightly smoothed with a Savitzky-Golay low-pass filter \citep{sg64}.

\begin{figure}[ht!]
\begin{center}
  \resizebox{\hsize}{!}{\includegraphics[]{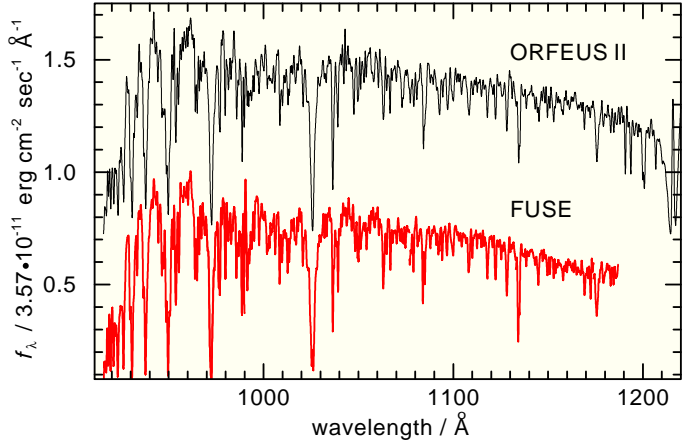}}
  \caption[]{Comparison of ORFEUS (shifted by 0.7 to the top) 
             and FUSE observations of \lb.  
            }
  \label{fig:ORFEUS}
\end{center}
\end{figure}

The FUSE spectrum exhibits a strong contamination by interstellar line absorption \sA{fig:ISM_MOD}.
To identify weak photospheric lines in the spectrum, we employed the program \owens\ 
\citep[cf\@.][]{Lemoine02, Hebrard02}. With \owens, we can simulate interstellar clouds with
individual parameters such as, e.g., radial velocity, column density in the line of sight, temperature of the
gas, and microturbulence velocity.
A large number of ions were taken into account, e.g\@. \Ion{H}{1}, \Ion{C}{2},
\Ion{C}{3}, \Ion{N}{1}, \Ion{N}{2}, \Ion{N}{3}, \Ion{O}{1}, \Ion{Si}{2}, \Ion{Ar}{1}, \Ion{Fe}{2}, and the $\mathrm{H}_2$ molecule
($J=0,1,2,3,4$). The continuum is well matched and most of
the absorption lines are well reproduced by the combined spectrum \sA{fig:ISM_MOD}.
The stellar spectrum is calculated with Kurucz' LIN lines (cf\@. \se{sect:analysis}) and, thus, the strong 
absorption feature at 979\,\AA\ (not observed) is most likely due to uncertain wavelengths in these lists.

\begin{figure}[ht!]
\begin{center}
  \resizebox{\hsize}{!}{\includegraphics[]{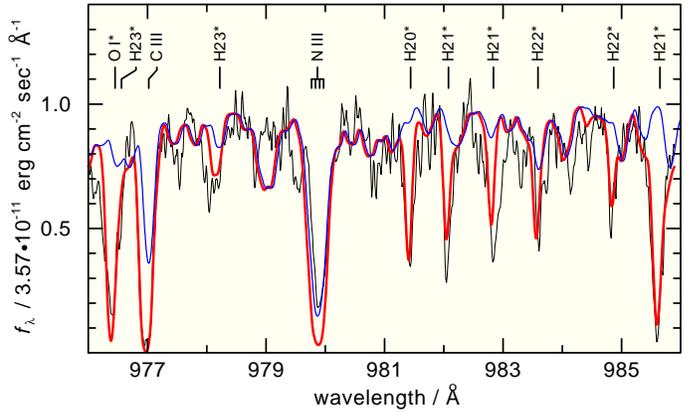}}
  \caption[]{Comparison of the synthetic stellar spectrum (thin line, blue) and the combined synthetic stellar + ISM spectrum
             (thick, red) with a section of the FUSE observation of \lb. 
             Interstellar lines are labeled with an asterisk.  
            }
  \label{fig:ISM_MOD}
\end{center}
\end{figure}

\citet{Rauch00} determined an interstellar neutral hydrogen column density of 
$n_\mathrm{H\,I}=2\cdot 10^{20}\,\mathrm{cm^{-2}}$ from Ly\,$\alpha$ (IUE observation).
We measured the same value from Ly\,$\beta$ in the FUSE observation \sA{fig:nH}.
The inner line core appears much too broad at \mbox{$n_\mathrm{H\,I}=4\cdot 10^{20}\,\mathrm{cm^{-2}}$}.

\begin{figure}[ht!]
  \resizebox{\hsize}{!}{\includegraphics[]{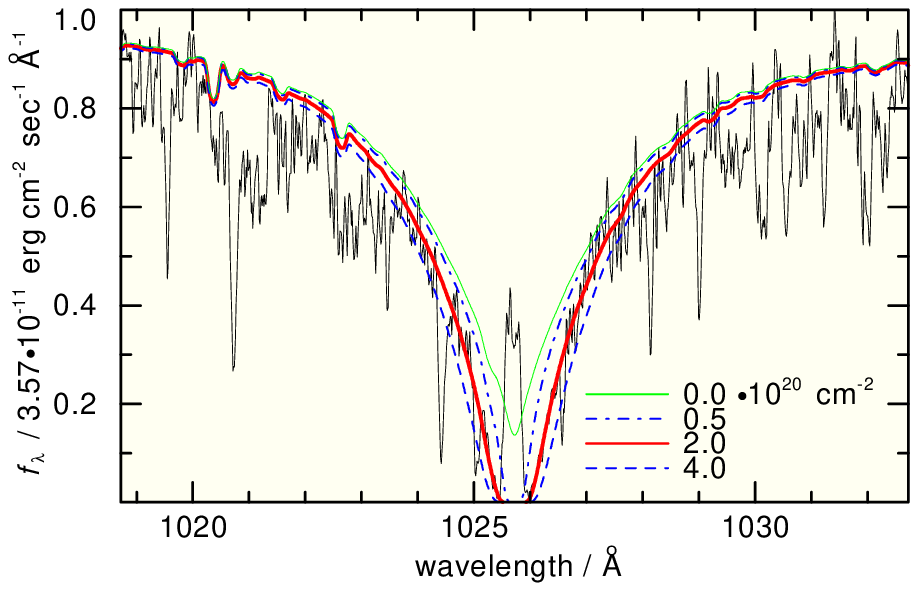}}
  \caption[]{Comparison of our synthetic Ly\,$\beta$ line profile (\Teffw{42}, \loggw{5.3}) at different $n_\mathrm{H\,I}$
             with the FUSE observation. The best fit is achieved at 
             \mbox{$n_\mathrm{H\,I}=2\cdot 10^{20}\,\mathrm{cm^{-20}}$}.
            }
  \label{fig:nH}
\end{figure}

In order to determine the interstellar reddening, we used FUSE and IUE observations \sA{fig:redd}.
We find that the continuum slope is best reproduced with $E_\mathrm{B-V}=0.01\pm 0.01$ using the
Galactic reddening law of \citet{Seaton79}. At this low  $E_\mathrm{B-V}$, we arrive at the same
result if the LMC reddening law of \citet{h83} is used. 
This value is significantly lower than $E_\mathrm{B-V}=0.0526$
\citep[cf\@.][]{Rauch00}, calculated from $n_\mathrm{H\,I}$ using an approximate formula given by 
\citet{GroenewegenLamers89}.

The ISM line absorption has been neglected in this particular determination.
Its consideration would decrease the flux level of our synthetic energy distribution at higher
energies and, thus, the derived reddening would be even lower. Since there are uncertainties, such as
in the location of the observed stellar continuum-flux level (difficult to determine in the presence 
of strong ISM absorption) and the validity of the applied reddening law, and the reddening is
relatively small, we decided to adopt $E_\mathrm{B-V}=0.00$ for our analysis without loss of generality.
Our determination of photospheric properties is not significantly affected by this assumption.

\ab{fig:spectrum} shows that -- even considering ISM line absorption -- an additional,
presently still unexplained factor, linearly increasing towards longer wavelengths,
appears necessary to achieve a good fit to the continuum in the FUSE wavelength range.

\begin{figure}[ht!]
  \resizebox{\hsize}{!}{\includegraphics[]{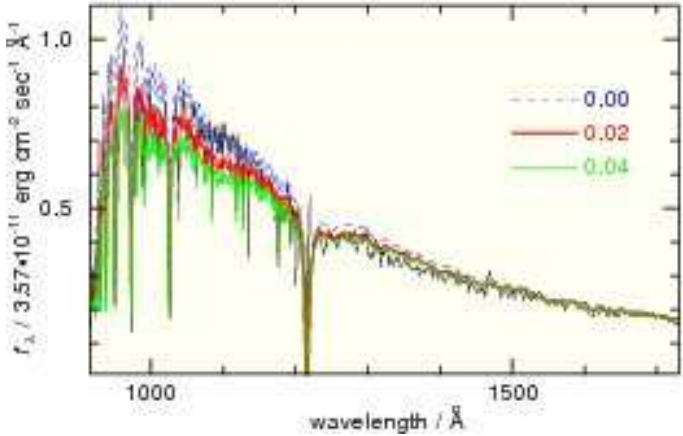}}
   \caption[]{Comparison of our synthetic spectrum calculated with Kurucz' LIN lines (cf\@. \se{sect:analysis})
              with the 
              FUSE ($\lambda < 1187\,\AA$),
              ORFEUS, and 
              IUE  ($\lambda > 1220\,\AA$, SWP04887LL retrieved from INES) 
              observations at three different $E_\mathrm{B-V}$.
              The IUE observation is scaled by a factor of 1.06 to match flux calibrations
              of other UV satellites \citep[see, e.g.,][]{Kruk97}.
              The synthetic spectra are normalized to match the flux level of the IUE observation at
              $\lambda=1700\,\AA$. 
              For clarity, at $\lambda < 1220\,\AA$ the synthetic spectra and the observation 
              are convolved with a Gaussian of 1\,\AA\ (FWHM). 
              At $\lambda > 1220\,\AA$ the synthetic spectra
              are convolved with a Gaussian of 7\,\AA\ (FWHM) to simulate IUE's low resolution.
             }
  \label{fig:redd}
\end{figure}

\begin{figure*}[ht!]
\begin{center}
  \resizebox{\hsize}{!}{\includegraphics[]{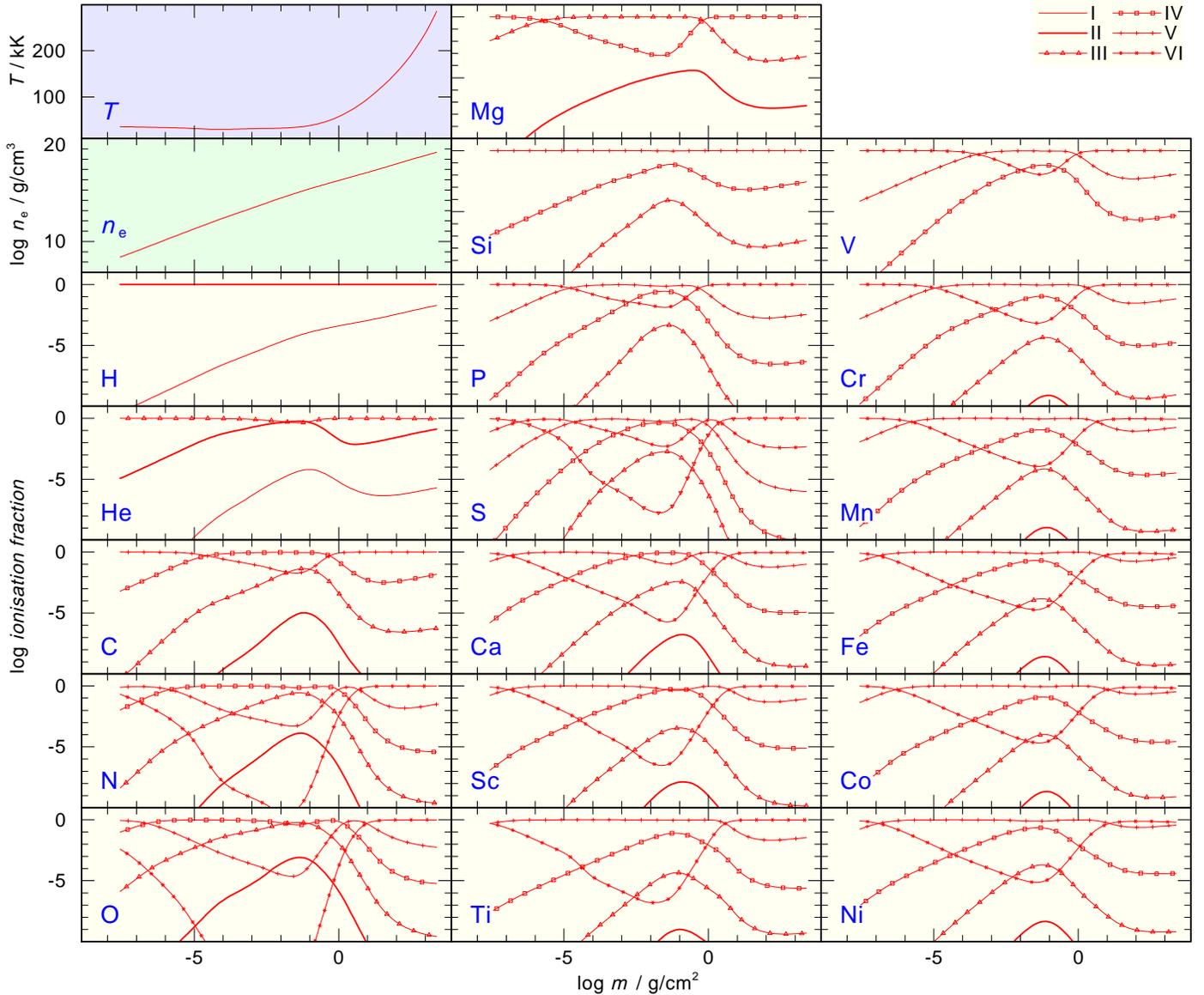}}
  \caption[]{Temperature and density structure, as well as ionization fractions of the elements
             that are considered in our atmosphere model of \lb.
            }
  \label{fig:fractions}
\end{center}
\end{figure*}

\section{Atomic data and modeling}
\label{sect:NLTE}

The model atmosphere of the primary of \lb\, was calculated with the T\"ubingen Model Atmosphere Package \tmap.
Details about \tmap\ are given by \citet{Werner03} and \citet{rauch03}.

In addition to H and He, the light metals C, N, O, Mg, Si, P, and S, as well as the elements of the iron group (Ca -- Ni), 
were considered. In our final model atmosphere, 567 levels are treated in NLTE with
724 individual lines and 801 superlines.
The statistics of our final model atmosphere are summarized in
\ta{tab:statistics}. The model atoms are based on atomic data taken from the databases of
NIST\footnote{\texttt{http://physics.nist.gov}} (National Institute of Standards and Technology), 
Opacity Project\footnote{\texttt{http://vizier.u-strasbg.fr/topbase/topbase.html}} \citep{Seaton94}, 
and Iron Project\footnote{\texttt{http://vizier.u-strasbg.fr/tipbase/tipbase.html}} \citep{Hummer93}. 
In the case of the iron-group elements, energy levels and oscillator strengths stem from
Kurucz' line lists \citep[][see \se{sect:analysis} for details]{Kurucz91}.

\begin{table}[ht!]
\caption[]{Statistics of the elements considered in our final model.} 
\label{tab:statistics}
\begin {tabular} {lrrrlrrrr} 
\hline
\hline
\noalign{\smallskip}
ion         & NL &  L & RBB & ion         & NL& RBB & $\mathrm{RBB_{st}}$\\
\hline
\noalign{\smallskip}
\Ion{H}{1}  & 15 &  1 & 105 & \Ion{Ca}{2} & 7 &  21 &         890 \\
\Ion{H}{2}  &  1 &  0 &   0 & \Ion{Ca}{3} & 7 &  25 &     10\,241 \\
\Ion{He}{1} & 29 & 15 &  61 & \Ion{Ca}{4} & 7 &  22 &     20\,291 \\
\Ion{He}{2} & 14 & 18 &  91 & \Ion{Ca}{5} & 7 &  24 &    141\,956 \\
\Ion{He}{3} &  1 &  0 &   0 & \Ion{Ca}{6} & 1 &   0 &           0 \\ 
\Ion{C}{2}  &  1 & 41 &   0 & \Ion{Sc}{2} & 7 &  26 &     36\,087 \\
\Ion{C}{3}  & 13 & 54 &  19 & \Ion{Sc}{3} & 7 &  20 &         675 \\
\Ion{C}{4}  & 14 & 44 &  35 & \Ion{Sc}{4} & 7 &  26 &     15\,024 \\
\Ion{C}{5}  &  1 &  0 &   0 & \Ion{Sc}{5} & 7 &  24 &     65\,994 \\
\Ion{N}{2}  &  1 & 21 &   0 & \Ion{Sc}{6} & 1 &   0 &           0 \\
\Ion{N}{3}  & 13 & 32 &  21 & \Ion{Ti}{2} & 7 &  26 &    154\,681 \\
\Ion{N}{4}  & 34 & 60 & 108 & \Ion{Ti}{3} & 7 &  24 &     21\,722 \\
\Ion{N}{5}  & 14 & 48 &  35 & \Ion{Ti}{4} & 7 &  24 &      1\,000 \\
\Ion{N}{6}  &  1 &  0 &   0 & \Ion{Ti}{5} & 7 &  24 &     26\,654 \\
\Ion{O}{2}  &  2 &  3 &   0 & \Ion{Ti}{6} & 1 &   0 &           0 \\
\Ion{O}{3}  & 19 & 35 &  33 & \Ion{V}{4}  & 7 &  24 &     37\,130 \\
\Ion{O}{4}  & 11 & 58 &  19 & \Ion{V}{5}  & 7 &  27 &      2\,123 \\
\Ion{O}{5}  &  5 &  7 &   3 & \Ion{V}{6}  & 1 &   0 &           0 \\
\Ion{O}{6}  &  1 &  0 &   0 & \Ion{Cr}{2} & 6 &  18 &    303\,129 \\
\Ion{Mg}{1} &  5 &  1 &   3 & \Ion{Cr}{3} & 7 &  25 &    523\,586 \\
\Ion{Mg}{2} & 14 & 19 &  34 & \Ion{Cr}{4} & 7 &  25 &    234\,170 \\
\Ion{Mg}{3} & 15 & 20 &  20 & \Ion{Cr}{5} & 7 &  24 &     43\,860 \\ 
\Ion{Mg}{4} &  1 &  0 &   0 & \Ion{Cr}{6} & 1 &   0 &           0 \\
\Ion{Si}{3} &  6 &  6 &   4 & \Ion{Mn}{2} & 6 &  20 &     68\,101 \\
\Ion{Si}{4} & 16 &  7 &  44 & \Ion{Mn}{3} & 7 &  22 &    671\,822 \\
\Ion{Si}{5} &  1 &  0 &   0 & \Ion{Mn}{4} & 7 &  25 &    719\,387 \\
\Ion{P}{3}  &  3 &  7 &   0 & \Ion{Mn}{5} & 7 &  25 &    285\,376 \\
\Ion{P}{4}  & 15 & 36 &   9 & \Ion{Mn}{6} & 1 &   0 &           0 \\
\Ion{P}{5}  & 18 &  7 &  12 & \Ion{Fe}{2} & 6 &  20 &    218\,490 \\
\Ion{P}{6}  &  1 &  0 &   0 & \Ion{Fe}{3} & 7 &  25 &    301\,981 \\
\Ion{S}{3}  &  1 &  9 &   0 & \Ion{Fe}{4} & 7 &  25 & 1\,027\,793 \\
\Ion{S}{4}  &  6 &  9 &   4 & \Ion{Fe}{5} & 7 &  25 &    793\,718 \\ 
\Ion{S}{5}  & 14 &  5 &  16 & \Ion{Fe}{6} & 1 &   0 &           0 \\
\Ion{S}{6}  & 18 &  7 &  48 & \Ion{Co}{2} & 6 &  18 &    244\,873 \\
\Ion{S}{7}  &  1 &  0 &   0 & \Ion{Co}{3} & 7 &  25 &    679\,280 \\
            &    &    &     & \Ion{Co}{4} & 7 &  25 &    552\,916 \\
            &    &    &     & \Ion{Co}{5} & 7 &  25 & 1\,469\,717 \\
            &    &    &     & \Ion{Co}{6} & 1 &   0 &           0 \\
            &    &    &     & \Ion{Ni}{2} & 6 &  18 &     96\,647 \\
            &    &    &     & \Ion{Ni}{3} & 7 &  22 &    418\,248 \\
            &    &    &     & \Ion{Ni}{4} & 7 &  25 &    949\,506 \\
            &    &    &     & \Ion{Ni}{5} & 7 &  27 & 1\,006\,189 \\
            &    &    &     & \Ion{Ni}{6} & 1 &   0 &           0 \\
\hline
\noalign{\smallskip}
\end{tabular}
For iron group elements, $\mathrm{RBB_{st}}$ denotes the number of individual lines which 
were summed up to so-called superlines \citep[see, e.g.,][]{rauch03}.
\end{table}

We adopted the parameters determined by \citet{Rauch00} (\Teffw{42}, \loggw{5.21}, and abundances) for our
first models. 
For additional species (Ca, Sc, Ti, V, Cr, Mn, Co), we assumed solar abundances \citep{asplund05}. 
\ab{fig:fractions} shows the ionization fractions of all elements in the model atmosphere. 
The ionization stages {\sc iii} and {\sc iv} dominate in the line-forming regions 
(\mbox{$\log m\approx-1$}) for the light metals, whereas {\sc iv} and {\sc v} are most populated in the case of the 
iron-group elements. 

A first test calculation was performed to check whether we can reproduce the results of
\citet{Rauch00} with our more elaborated models (e.g\@. we consider for Fe and Ni opacities of the
ionization stages {\sc ii} -- {\sc vi} in contrast to {\sc iv} -- {\sc ix}). \aba{fig:IUE} demonstrates
that the IUE near-UV observation is reproduced well with adjusted Si and Ni abundances (about a factor of two lower
and a factor of two higher, respectively, compared to the previous results).

\begin{figure}[ht]
  \resizebox{\hsize}{!}{\includegraphics[]{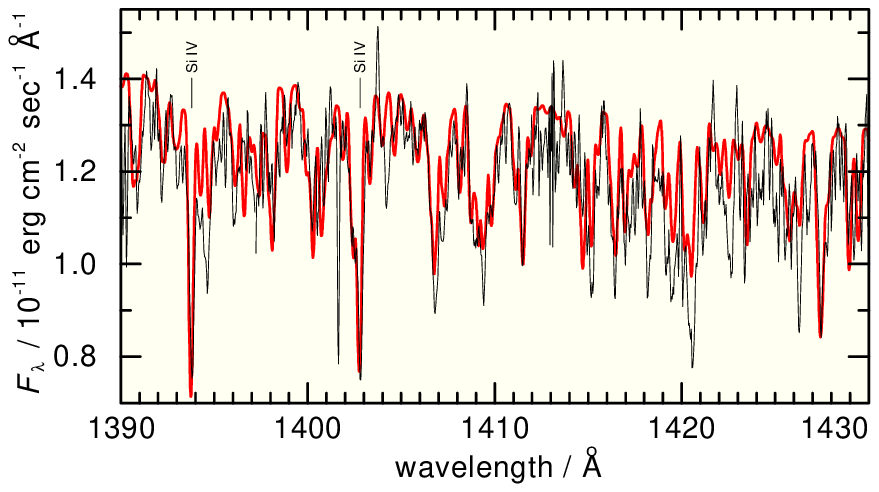}}
  \caption[]{Comparison of our synthetic spectrum (thick, red line, \Teffw{42}, \loggw{5.3}, 
             abundances see \ab{fig:abundances})
             with the IUE observation \citep[thin line, SWP\,17822, cf\@.][his Fig.\,14]{Rauch00}
             of \lb.}
  \label{fig:IUE}
\end{figure}

For a combined synthetic spectrum that includes both 
the synthetic stellar as well as the ISM spectrum, we first normalized the ISM spectrum calculated by \owens\ 
and multiplied it by the \tmap\ spectrum, which has been convolved with a rotational profile 
(35\,km/sec unless otherwise noted, cf\@. \se{sect:vrot}) 
before. Finally, the combined spectrum was convolved with a Gaussian of 0.05\,\AA\ (FWHM)
to match the instrument's resolution.

The apparent continuum slope in the FUSE spectrum changes at approximately 1100\,\AA, becoming
noticeably shallower on the shorter-wavelength side of this point. To facilitate detailed comparison
of the small-scale structure in the models with the data, the models shown in \ab{fig:spectrum} have been normalized
by a factor that increases linearly from $5.6 \cdot 10^{18}$ at 910\,\AA\ to $7.9 \cdot 10^{18}$ at 1190\,\AA.

\subsection{Identification of photospheric lines}
\label{sect:lines}

In the FUSE observation of \lb\ \sA{fig:spectrum}, several photospheric absorption lines are prominent and isolated 
from ISM absorption lines. In \ta{tab:photospheric}, we summarize the identified photospheric lines.
For the first time, we have identified 
\Ion{P}{4} and \Ion{P}{5} \sA{fig:phosphorus},
as well as
\Ion{S}{4} and \Ion{S}{6} \sA{fig:sulfur},
lines in the spectrum of \lb.

\begin{figure*}[ht!]
  \resizebox{0.98\hsize}{!}{\includegraphics[]{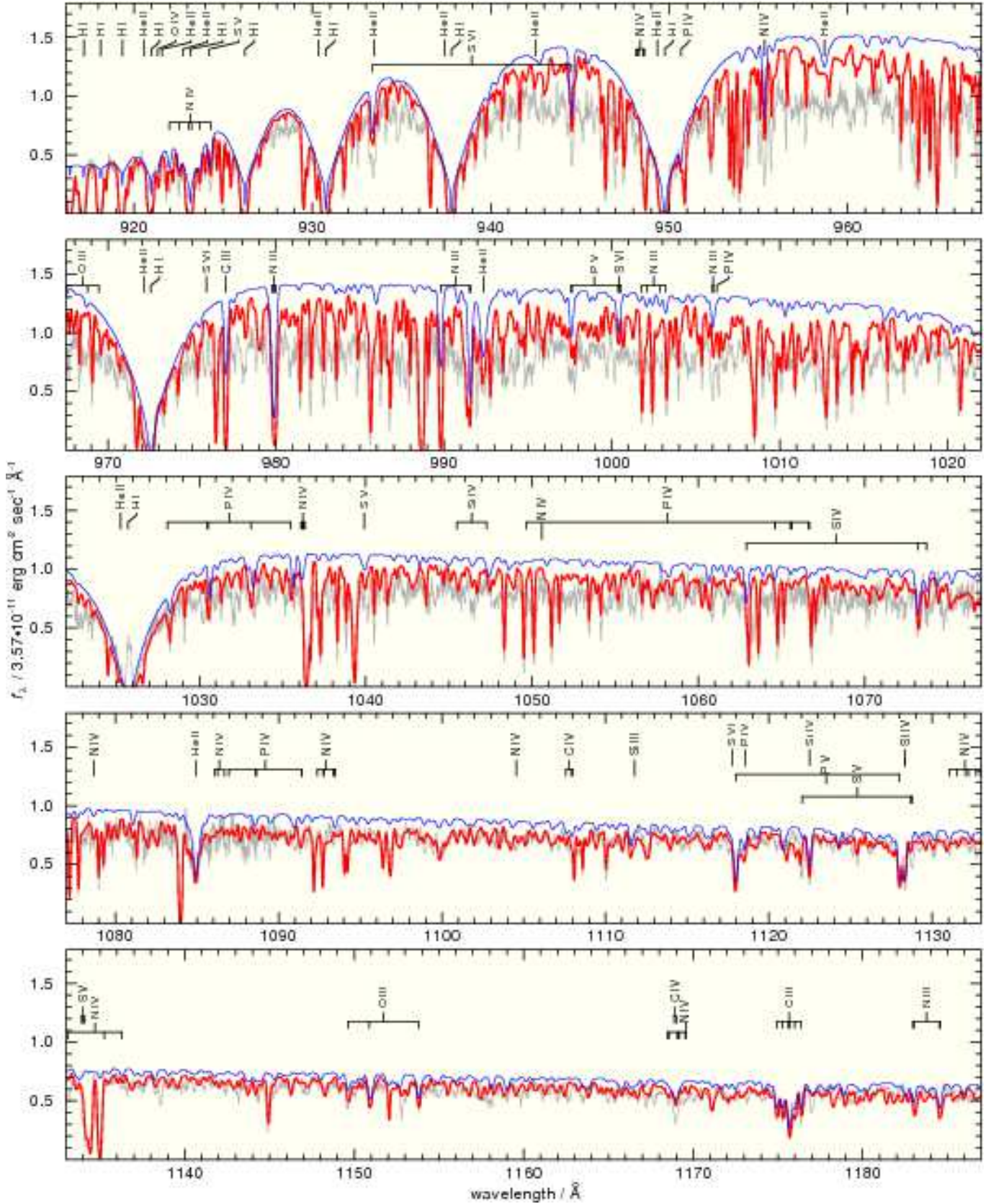}}
  \caption[]{Comparison of the 
             synthetic stellar spectrum of our final HHeCNOMgSiPS+CaScTiVCrMnFeCoNi NLTE model (\Teffw{42}, \loggw{5.3}) 
             (blue, thin line, Kurucz' POS lines)
             and the 
             synthetic stellar + ISM spectrum (thick red, Kurucz LIN lines) 
             with the FUSE observation. 
             The observation is shifted to 
             match the rest wavelengths of the photospheric lines marked at the top of the panels.
             The synthetic spectra are normalized to match the observed flux level with a factor
             linearly increasing from $5.6\cdot 10^{18}$ to $7.9\cdot 10^{18}$ within 910\,\AA\ -- 1190\,\AA.
            }
  \label{fig:spectrum}
\end{figure*}

\begin{table}[htb!]
\caption[]{Identified photospheric lines in the FUSE observation of \lb,
           with only \Ion{P}{4} $3\rm{p}^{~}\,^3\rm{P}^o$ - $3\rm{p}^2\,^3\rm{P}^{~}$ lines identified,
           and other \Ion{P}{4} line identifications uncertain. }
\label{tab:photospheric}
\begin{tabular}{r@{.}l@{ -- }r@{.}llr@{ -- }l} 
\hline
\hline
\noalign{\smallskip}
\multicolumn{4}{c}{Wavelength / \AA} &\hspace{5pt} Ion & \multicolumn{2}{c}{Transition} \\
\hline
\noalign{\smallskip}
 918&\multicolumn{3}{l}{\hspace{-6pt}50} &\hspace{5pt} \Ion{P}{4}  &\hspace{5pt} $3\rm{p}^2\,^1\rm{D}^{~}$   & $3\rm{d}^{~}\,^1\rm{D}^o$   \vspace{-1.7pt}\\
 920&\multicolumn{3}{l}{\hspace{-6pt}55} &\hspace{5pt} \Ion{He}{2} &\hspace{5pt} 2 & 20                                                    \vspace{-1.7pt}\\
 920&\multicolumn{3}{l}{\hspace{-6pt}96} &\hspace{5pt} \Ion{H}{1}  &\hspace{5pt} 1 & 10                                                    \vspace{-1.7pt}\\
 921&\multicolumn{3}{l}{\hspace{-6pt}32} &\hspace{5pt} \Ion{O}{4}  &\hspace{5pt} $2\rm{p}^2\,^2\rm{P}^{~}$   & $2\rm{p}^3\,^2\rm{P}^o$     \vspace{-1.7pt}\\
 921&\multicolumn{3}{l}{\hspace{-6pt}56} &\hspace{5pt} \Ion{He}{2} &\hspace{5pt} 2 & 19                                                    \vspace{-1.7pt}\\
 921&99 & 924&28 &\hspace{5pt}                         \Ion{N}{4}  &\hspace{5pt} $2\rm{p}^{~}\,^3\rm{P}^o$   & $2\rm{p}^2\,^3\rm{P}^{~}$   \vspace{-1.7pt}\\
 922&\multicolumn{3}{l}{\hspace{-6pt}74} &\hspace{5pt} \Ion{He}{2} &\hspace{5pt} 2 & 18                                                    \vspace{-1.7pt}\\
 923&\multicolumn{3}{l}{\hspace{-6pt}15} &\hspace{5pt} \Ion{H}{1}  &\hspace{5pt} 1 & 9                                                     \vspace{-1.7pt}\\
 924&\multicolumn{3}{l}{\hspace{-6pt}13} &\hspace{5pt} \Ion{He}{2} &\hspace{5pt} 2 & 17                                                    \vspace{-1.7pt}\\
 924&\multicolumn{3}{l}{\hspace{-6pt}22} &\hspace{5pt} \Ion{S}{5}  &\hspace{5pt} $3\rm{p}^{~}\,^1\rm{P}^o$   & $3\rm{p}^2\,^1\rm{S}^{~}$   \vspace{-1.7pt}\\
 925&\multicolumn{3}{l}{\hspace{-6pt}81} &\hspace{5pt} \Ion{He}{2} &\hspace{5pt} 2 & 16                                                    \vspace{-1.7pt}\\
 926&\multicolumn{3}{l}{\hspace{-6pt}23} &\hspace{5pt} \Ion{H}{1}  &\hspace{5pt} 1 & 8                                                     \vspace{-1.7pt}\\
 927&\multicolumn{3}{l}{\hspace{-6pt}85} &\hspace{5pt} \Ion{He}{2} &\hspace{5pt} 2 & 15                                                    \vspace{-1.7pt}\\
 930&\multicolumn{3}{l}{\hspace{-6pt}34} &\hspace{5pt} \Ion{He}{2} &\hspace{5pt} 2 & 14                                                    \vspace{-1.7pt}\\
 930&\multicolumn{3}{l}{\hspace{-6pt}75} &\hspace{5pt} \Ion{H}{1}  &\hspace{5pt} 1 & 7                                                     \vspace{-1.7pt}\\
 933&38 & 944&52 &\hspace{5pt}                         \Ion{S}{6}  &\hspace{5pt} $3\rm{s}^{~}\,^2\rm{S}^{~}$ & $3\rm{p}^{~}\,^2\rm{P}^o$   \vspace{-1.7pt}\\
 933&\multicolumn{3}{l}{\hspace{-6pt}45} &\hspace{5pt} \Ion{He}{2} &\hspace{5pt} 2 & 13                                                    \vspace{-1.7pt}\\
 937&\multicolumn{3}{l}{\hspace{-6pt}39} &\hspace{5pt} \Ion{He}{2} &\hspace{5pt} 2 & 12                                                    \vspace{-1.7pt}\\
 937&\multicolumn{3}{l}{\hspace{-6pt}80} &\hspace{5pt} \Ion{H}{1}  &\hspace{5pt} 1 & 6                                                     \vspace{-1.7pt}\\
 942&\multicolumn{3}{l}{\hspace{-6pt}51} &\hspace{5pt} \Ion{He}{2} &\hspace{5pt} 2 & 11                                                    \vspace{-1.7pt}\\
 948&09 & 948&21 &\hspace{5pt}                         \Ion{C}{4}  &\hspace{5pt} $3\rm{s}^{~}\,^2\rm{S}^{~}$ & $4\rm{p}^{~}\,^2\rm{P}^o$   \vspace{-1.7pt}\\
 948&15 & 948&61 &\hspace{5pt}                         \Ion{N}{4}  &\hspace{5pt} $3\rm{p}^{~}\,^3\rm{P}^o$   & $4\rm{d}^{~}\,^3\rm{D}^{~}$ \vspace{-1.7pt}\\
 949&\multicolumn{3}{l}{\hspace{-6pt}33} &\hspace{5pt} \Ion{He}{2} &\hspace{5pt} 2 & 10                                                    \vspace{-1.7pt}\\
 949&\multicolumn{3}{l}{\hspace{-6pt}74} &\hspace{5pt} \Ion{H}{1}  &\hspace{5pt} 1 & 5                                                     \vspace{-1.7pt}\\
 950&\multicolumn{3}{l}{\hspace{-6pt}66} &\hspace{5pt} \Ion{P}{4}  &\hspace{5pt} $3\rm{s}^2\,^1\rm{S}^{~}$   & $3\rm{p}^{~}\,^1\rm{P}^o$   \vspace{-1.7pt}\\
 955&\multicolumn{3}{l}{\hspace{-6pt}34} &\hspace{5pt} \Ion{N}{4}  &\hspace{5pt} $2\rm{p}^{~}\,^1\rm{P}^o$   & $2\rm{p}^2\,^1\rm{S}^{~}$   \vspace{-1.7pt}\\
 958&\multicolumn{3}{l}{\hspace{-6pt}70} &\hspace{5pt} \Ion{He}{2} &\hspace{5pt} 2 & 9                                                     \vspace{-1.7pt}\\
 967&66 & 968&18 &\hspace{5pt}                         \Ion{O}{3}  &\hspace{5pt} $2\rm{p}^3\,^1\rm{P}^o$     & $3\rm{p}^{~}\,^1\rm{S}^{~}$ \vspace{-1.7pt}\\
 972&\multicolumn{3}{l}{\hspace{-6pt}11} &\hspace{5pt} \Ion{He}{2} &\hspace{5pt} 2 & 8                                                     \vspace{-1.7pt}\\
 972&\multicolumn{3}{l}{\hspace{-6pt}53} &\hspace{5pt} \Ion{H}{1}  &\hspace{5pt} 1 & 4                                                     \vspace{-1.7pt}\\
 975&\multicolumn{3}{l}{\hspace{-6pt}84} &\hspace{5pt} \Ion{S}{6}  &\hspace{5pt} $4\rm{p}^{~}\,^2\rm{P}^o$   & $5\rm{s}^{~}\,^2\rm{S}^{~}$ \vspace{-1.7pt}\\
 977&\multicolumn{3}{l}{\hspace{-6pt}02} &\hspace{5pt} \Ion{C}{3}  &\hspace{5pt} $2\rm{s}^2\,^1\rm{S}^{~}$   & $2\rm{p}^{~}\,^1\rm{P}^o$   \vspace{-1.7pt}\\
 979&77 & 979&97 &\hspace{5pt}                         \Ion{N}{3}  &\hspace{5pt} $2\rm{p}^2\,^2\rm{D}^{~}$   & $2\rm{p}^3\,^2\rm{D}^o$     \vspace{-1.7pt}\\
 989&80 & 991&58 &\hspace{5pt}                         \Ion{N}{3}  &\hspace{5pt} $2\rm{p}^{~}\,^2\rm{P}^o$   & $2\rm{p}^2\,^2\rm{D}^{~}$   \vspace{-1.7pt}\\
 992&\multicolumn{3}{l}{\hspace{-6pt}36} &\hspace{5pt} \Ion{He}{2} &\hspace{5pt} 2 & 7                                                     \vspace{-1.7pt}\\
 997&54 & 1000&38 &\hspace{5pt}                        \Ion{P}{5}  &\hspace{5pt} $3\rm{d}^{~}\,^2\rm{D}^{~}$ & $4\rm{p}^{~}\,^2\rm{P}^o$   \vspace{-1.7pt}\\ 
1000&37 & 1000&54 &\hspace{5pt}                        \Ion{S}{6}  &\hspace{5pt} $4\rm{d}^{~}\,^2\rm{D}^{~}$ & $5\rm{f}^{~}\,^2\rm{F}^o$   \vspace{-1.7pt}\\
1001&75 & 1003&21 &\hspace{5pt}                        \Ion{N}{3}  &\hspace{5pt} $2\rm{p}^2\,^2\rm{P}^{~}$   & $3\rm{p}^{~}\,^2\rm{P}^o$   \vspace{-1.7pt}\\
1005&99 & 1006&03 &\hspace{5pt}                        \Ion{N}{3}  &\hspace{5pt} $2\rm{p}^2\,^2\rm{S}^{~}$   & $2\rm{p}^3\,^2\rm{P}^o$     \vspace{-1.7pt}\\
1006&\multicolumn{3}{l}{\hspace{-6pt}23} &\hspace{5pt} \Ion{P}{4}  &\hspace{5pt} $3\rm{p}^2\,^1\rm{D}^{~}$   & $4\rm{p}^{~}\,^1\rm{P}^o$   \vspace{-1.7pt}\\
1025&\multicolumn{3}{l}{\hspace{-6pt}27} &\hspace{5pt} \Ion{He}{2} &\hspace{5pt} 2 & 6                                                     \vspace{-1.7pt}\\
1025&\multicolumn{3}{l}{\hspace{-6pt}56} &\hspace{5pt} \Ion{P}{4}  &\hspace{5pt} $3\rm{p}^{~}\,^3\rm{P}^o$   & $3\rm{p}^2\,^3\rm{P}^{~}$   \vspace{-1.7pt}\\
1025&\multicolumn{3}{l}{\hspace{-6pt}72} &\hspace{5pt} \Ion{H}{1}  &\hspace{5pt} 1 & 3                                                     \vspace{-1.7pt}\\
1028&09 & 1035&52 &\hspace{5pt}                        \Ion{P}{4}  &\hspace{5pt} $3\rm{p}^{~}\,^3\rm{P}^o$   & $3\rm{p}^2\,^3\rm{P}^{~}$   \vspace{-1.7pt}\\
1039&\multicolumn{3}{l}{\hspace{-6pt}92} &\hspace{5pt} \Ion{S}{5}  &\hspace{5pt} $3\rm{d}^{~}\,^1\rm{D}^{~}$ & $3\rm{d}^{~}\,^1\rm{F}^o$   \vspace{-1.7pt}\\
1045&50 & 1047&27 &\hspace{5pt}                        \Ion{Si}{4} &\hspace{5pt} $4\rm{p}^{~}\,^2\rm{P}^o$   & $6\rm{d}^{~}\,^2\rm{D}^{~}$ \vspace{-1.7pt}\\
1049&\multicolumn{3}{l}{\hspace{-6pt}65} &\hspace{5pt} \Ion{P}{4}  &\hspace{5pt} $3\rm{d}^{~}\,^1\rm{D}^{~}$ & $3\rm{d}^{~}\,^1\rm{F}^o$   \vspace{-1.7pt}\\
1050&\multicolumn{3}{l}{\hspace{-6pt}60} &\hspace{5pt} \Ion{N}{4}  &\hspace{5pt} $3\rm{p}^{~}\,^1\rm{P}^o$   & $3\rm{p}^{~}\,^1\rm{D}^{~}$ \vspace{-1.7pt}\\
1062&\multicolumn{3}{l}{\hspace{-6pt}88} &\hspace{5pt} \Ion{S}{4}  &\hspace{5pt} $3\rm{p}^{~}\,^2\rm{P}^{o}$ & $3\rm{p}^{2}\,^2\rm{D}^{~}$ \vspace{-1.7pt}\\
1064&61 & 1066&64 &\hspace{5pt}                        \Ion{P}{4}  &\hspace{5pt} $3\rm{d}^{~}\,^3\rm{D}^{~}$ & $3\rm{d}^{~}\,^1\rm{F}^o$   \vspace{-1.7pt}\\
1073&19 & 1073&72 &\hspace{5pt}                        \Ion{S}{4}  &\hspace{5pt} $3\rm{p}^{~}\,^2\rm{P}^{o}$ & $3\rm{p}^{2}\,^2\rm{D}^{~}$ \vspace{-1.7pt}\\
1078&\multicolumn{3}{l}{\hspace{-6pt}71} &\hspace{5pt} \Ion{N}{4}  &\hspace{5pt} $3\rm{d}^{~}\,^1\rm{D}^{~}$ & $4\rm{f}^{~}\,^1\rm{F}^o$   \vspace{-1.7pt}\\
1084&\multicolumn{3}{l}{\hspace{-6pt}94} &\hspace{5pt} \Ion{He}{2} &\hspace{5pt} 2 & 5                                                     \vspace{-1.7pt}\\
1086&08 & 1086&69 &\hspace{5pt}                        \Ion{N}{4}  &\hspace{5pt} $3\rm{p}^{~}\,^3\rm{P}^o$ & $3\rm{p}^{~}\,^3\rm{S}^{~}$   \vspace{-1.7pt}\\
1086&93 & 1091&44 &\hspace{5pt}                        \Ion{P}{4}  &\hspace{5pt} $3\rm{d}^{~}\,^3\rm{D}^{~}$ & $3\rm{d}^{~}\,^3\rm{D}^o$   \vspace{-1.7pt}\\
1092&35 & 1093&48 &\hspace{5pt}                        \Ion{N}{4}  &\hspace{5pt} $3\rm{d}^{~}\,^3\rm{D}^{~}$ & $3\rm{d}^{~}\,^3\rm{P}^o$   \vspace{-1.7pt}\\
1097&32 & 1097&34 &\hspace{5pt}                        \Ion{C}{4}  &\hspace{5pt} $4\rm{s}^{~}\,^2\rm{S}^{~}$ & $8\rm{p}^{~}\,^2\rm{P}^o$   \vspace{-1.7pt}\\
1104&\multicolumn{3}{l}{\hspace{-6pt}53} &\hspace{5pt} \Ion{N}{4}  &\hspace{5pt} $3\rm{d}^{~}\,^3\rm{F}^o$   & $4\rm{p}^{~}\,^3\rm{D}^{~}$ \vspace{-1.7pt}\\
1107&59 & 1107&98 &\hspace{5pt}                        \Ion{C}{4}  &\hspace{5pt} $3\rm{p}^{~}\,^2\rm{P}^o$   & $4\rm{d}^{~}\,^2\rm{D}^{~}$ \vspace{-1.7pt}\\
1117&\multicolumn{3}{l}{\hspace{-6pt}76} &\hspace{5pt} \Ion{S}{6}  &\hspace{5pt} $4\rm{f}^{~}\,^2\rm{F}^o$   & $5\rm{g}^{~}\,^2\rm{G}^{~}$ \vspace{-1.7pt}\\
1117&\multicolumn{3}{l}{\hspace{-6pt}93} &\hspace{5pt} \Ion{N}{4}  &\hspace{5pt} $3\rm{d}^{~}\,^1\rm{D}^{~}$ & $3\rm{d}^{~}\,^1\rm{P}^o$   \vspace{-1.7pt}\\
1117&98 & 1128&01 &\hspace{5pt}                        \Ion{P}{5}  &\hspace{5pt} $3\rm{s}^{~}\,^2\rm{S}^{~}$ & $3\rm{p}^{~}\,^2\rm{P}^o$   \vspace{-1.7pt}\\
1118&\multicolumn{3}{l}{\hspace{-6pt}55} &\hspace{5pt} \Ion{P}{4}  &\hspace{5pt} $3\rm{p}^{~}\,^1\rm{P}^o$   & $3\rm{p}^2\,^1\rm{S}^{~}$   \vspace{-1.7pt}\\
1122&03 & 1134&09 &\hspace{5pt}                        \Ion{S}{5}  &\hspace{5pt} $3\rm{d}^{~}\,^3\rm{D}^{~}$ & $3\rm{d}^{~}\,^3\rm{F}^o$   \vspace{-1.7pt}\\
1122&49 & 1128&34 &\hspace{5pt}                        \Ion{Si}{4} &\hspace{5pt} $3\rm{p}^{~}\,^2\rm{P}^o$   & $3\rm{d}^{~}\,^2\rm{D}^{~}$ \vspace{-1.7pt}\\
1131&03 & 1132&94 &\hspace{5pt}                        \Ion{N}{4}  &\hspace{5pt} $3\rm{p}^{~}\,^3\rm{P}^o$   & $3\rm{p}^{~}\,^3\rm{P}^{~}$ \vspace{-1.7pt}\\
1133&12 & 1136&27 &\hspace{5pt}                        \Ion{N}{4}  &\hspace{5pt} $3\rm{s}^{~}\,^3\rm{S}^{~}$ & $3\rm{s}^{~}\,^3\rm{P}^o$   \vspace{-1.7pt}\\
1137&\multicolumn{3}{l}{\hspace{-6pt}28} &\hspace{5pt} \Ion{P}{4}  &\hspace{5pt} $3\rm{d}^{~}\,^1\rm{D}^{~}$ & $3\rm{d}^{~}\,^1\rm{P}^o$   \vspace{-1.7pt}\\
1149&63 & 1153&78 &\hspace{5pt}                        \Ion{O}{3}  &\hspace{5pt} $2\rm{p}^3\,^3\rm{S}^o$     & $2\rm{p}^4\,^3\rm{P}^{~}$   \vspace{-1.7pt}\\
1168&48 & 1169&57 &\hspace{5pt}                        \Ion{N}{4}  &\hspace{5pt} $3\rm{d}^{~}\,^3\rm{D}^{~}$ & $3\rm{d}^{~}\,^3\rm{D}^o$   \vspace{-1.7pt}\\
1168&86 & 1169&01 &\hspace{5pt}                        \Ion{C}{4}  &\hspace{5pt} $3\rm{d}^{~}\,^2\rm{D}^{~}$ & $4\rm{f}^{~}\,^2\rm{F}^o$   \vspace{-1.7pt}\\
1174&93 & 1176&37 &\hspace{5pt}                        \Ion{C}{3}  &\hspace{5pt} $2\rm{p}^{~}\,^3\rm{P}^o$   & $2\rm{p}^2\,^3\rm{P}^{~}$   \vspace{-1.7pt}\\
1182&97 & 1184&57 &\hspace{5pt}                        \Ion{N}{3}  &\hspace{5pt} $2\rm{p}^2\,^2\rm{P}^{~}$   & $2\rm{p}^3\,^2\rm{P}^o$     \vspace{-1.7pt}\\  
\hline
\end{tabular}
\end{table}

\begin{figure}[ht!]
\begin{center}
  \resizebox{\hsize}{!}{\includegraphics[]{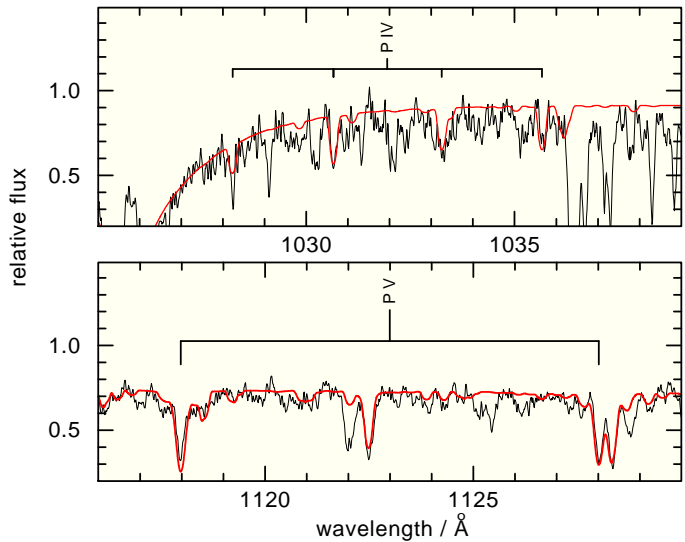}}
  \caption[]{Comparison of synthetic spectra (\Teffw{42}, \loggw{5.3}, P with solar abundance) around 
             \Ionww{P}{4}{1028.09, 1030.51, 1033.11, 1035.51} (top)
             and the \Ionww{P}{5}{1117.98, 1128.01} resonance doublet (bottom) 
             with the FUSE observation.  
            }
  \label{fig:phosphorus}
\end{center}
\end{figure} 

\begin{figure}[ht!]
\begin{center}
  \resizebox{\hsize}{!}{\includegraphics[]{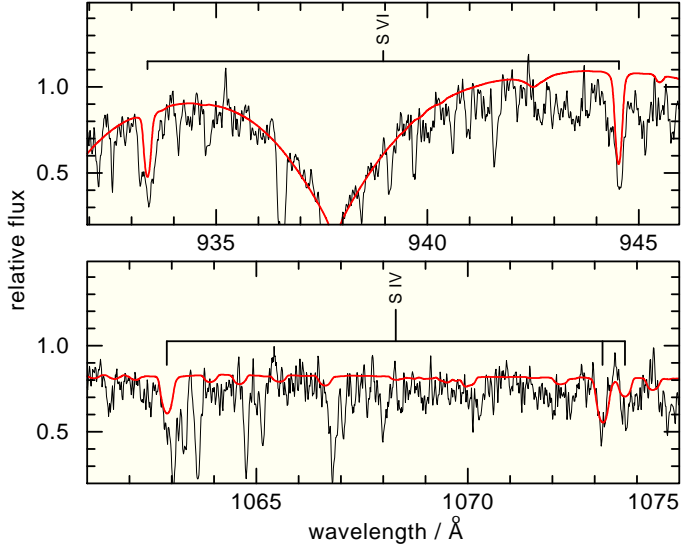}}
  \caption[]{Comparison of synthetic spectra (\Teffw{42}, \loggw{5.3}, sulfur with 0.01 times solar abundance) around 
             the \Ionww{S}{6}{933.38, 944.52} resonance doublet (top)
             and \Ionww{S}{4}{1062.88, 1073.19, 1073.72} (bottom) 
             with the FUSE observation.
            }
  \label{fig:sulfur}
\end{center}
\end{figure}

\citet{Rauch00} identified a large number of strong absorption lines of 
Fe\,{\sc iv}-{\sc v} and Ni\,{\sc iv}-{\sc v} in IUE observations of \lb. 
In the FUSE observation we cannot identify isolated lines of the iron-group elements because only weak lines are
located in this wavelength range, and the strong ISM line absorption hampers the search for such weak lines.

\section{Spectral analysis}
\label{sect:analysis}

A detailed spectral analysis of optical and near-UV observations \lb\ had been performed by \citet{Rauch00}.
We considered additional species \sT{tab:statistics} in order to model the far-UV spectrum reliably.
Although no iron-group lines were identified in the FUSE observation, we included all iron-group
elements with individual model atoms \sT{tab:statistics} using the Kurucz line lists \citep{Kurucz91}.
The model atoms and the respective atomic data files were constructed in a statistical approach
(introducing ``superlevels'' and ``superlines'') with the
program \ironic\ \citep[Iron Opacity Interface,][]{Deetjen99, rauch03}.
For the model-atmosphere calculation, we considered all lines (so-called LIN lists that include laboratory 
measured, as well as theoretically calculated lines) to simulate the total opacity correctly. 
\aba{fig:spectrum} shows the good agreement of our final synthetic ``LIN'' spectrum with the FUSE observation.
In a close inspection, e.g\@. around $\lambda\,960\,\mathrm{\AA}$ or $\lambda\,1015\,\mathrm{\AA}$, 
opacity appears to be missing. This might be the result of partly uncertain ``LIN''-line wavelengths.

For a detailed comparison with the observations, e.g\@. for identification and abundance determination,
we have to restrict ourselves to lines measured in the laboratory (using POS lists -- about 10\,\% of the lines in the LIN lists;
\ab{fig:spectrum}).

\subsection{Rotational velocity}
\label{sect:vrot}

The spectral analysis of \citet{Rauch00} was hampered by an uncertain rotational velocity.
The strongest photospheric line feature in the FUSE observation of \lb\ \sA{fig:spectrum}, \Ionww{C}{3}{1174 - 1177}, 
is well-suited to measuring the rotational velocity \sA{fig:vrot}. 
We determined $v_\mathrm{rot}=35\pm 5\,\mathrm{km/sec}$. 
This agrees with $v_\mathrm{rot}=34\pm 10\,\mathrm{km/sec}$, which was calculated by \citet{Rauch00} in an 
attempt to numerically eliminate the effects of orbital smearing from their spectra with relatively long 
exposure times. However, \citet{Rauch00} and \citet{RauchWerner03} assumed this $v_\mathrm{rot}$ was uncertain and used
$v_\mathrm{rot}=45\,\mathrm{km/sec}$ in their analyses. 
Our result now agrees with the $v_\mathrm{rot}=34.7-38.7\,\mathrm{km/sec}$ 
calculated under the assumption of bound rotation from the primary's radius $R_1 = 0.179 - 0.200\,\mathrm{R_\odot}$
given by \citet{Hilditch03}.

\begin{figure}[ht]
  \resizebox{\hsize}{!}{\includegraphics[]{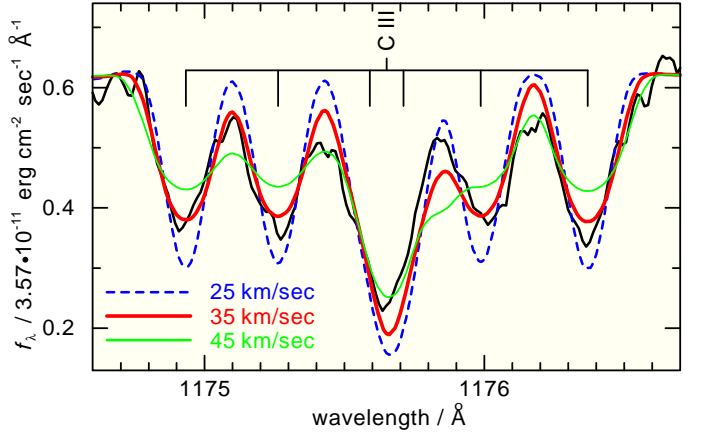}}
  \caption[]{Comparison of our synthetic spectrum (\Teffw{42}, \loggw{5.3}) with the FUSE observation 
             around \Ionww{C}{3}{1174.93-1176.37}.
             The synthetic spectrum is convolved with rotational profiles 
             ($v_\mathrm{rot}=25, 35, 45\,\mathrm{km/sec}$).
             The best fit is achieved at $v_\mathrm{rot}=35\,\mathrm{km/sec}$.
            }
  \label{fig:vrot}
\end{figure}

From $v_\mathrm{rot}=35\pm 5\,\mathrm{km/sec}$ and an orbital period of $P=22\,597.033\,\mathrm{sec}$ \citep{Kilkenny00}, 
we can calculate a stellar radius of $r_1=0.181\pm 0.025\,\mathrm{R_\odot}$. 
\citet{RauchWerner03} measured a radial-velocity amplitude of $A_1 = 39.19\pm 0.05\,\mathrm{km/sec}$ and 
calculated $a_1 = 0.2025\pm 0.0019\mathrm{R_\odot}$. 
With $a_1 = 0.2025\mathrm{R_\odot}$ and $r_1=0.181\,\mathrm{R_\odot}$, we can calculate $v_\mathrm{rot}=31\,\mathrm{km/sec}$,
which is in good agreement. Thus, it is most likely that the rotation of \lb\ is bound.

\subsection{Effective temperature}
\label{sect:Teff}

\begin{figure}[ht]
  \resizebox{\hsize}{!}{\includegraphics[]{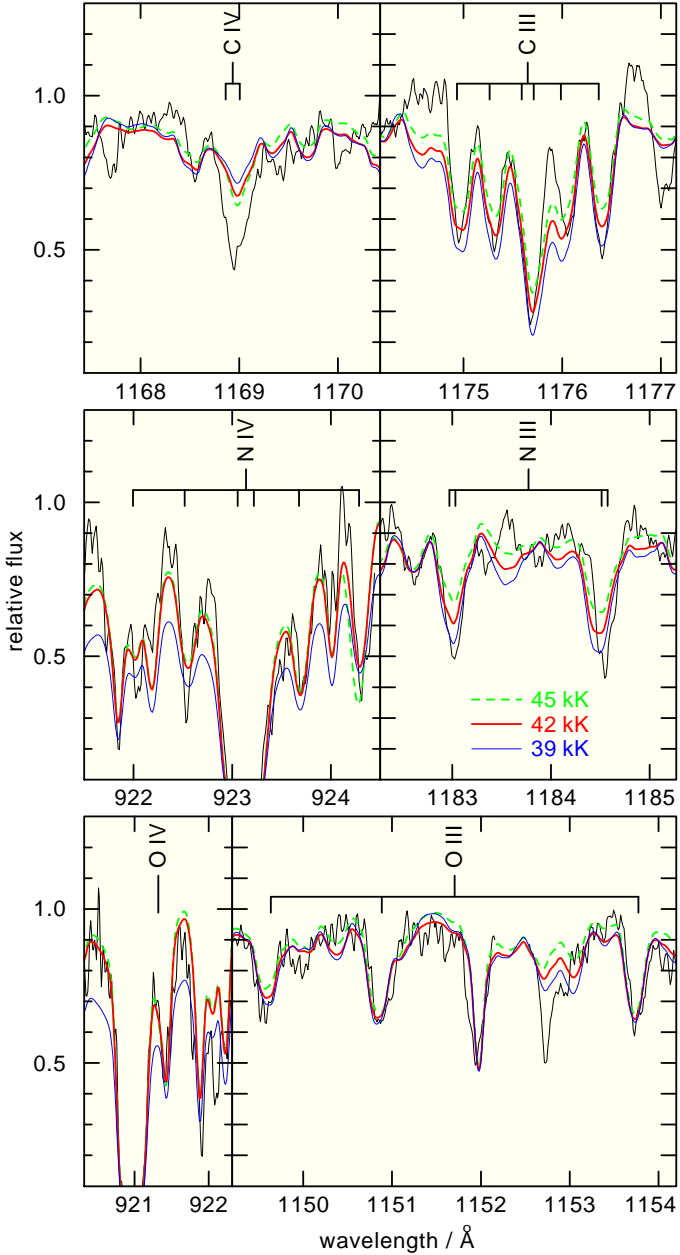}}
  \caption[]{Comparison of synthetic spectra (\loggw{5.3}) around selected
             \Ion{C}{4} and \Ion{C}{3} (top panel), 
             \Ion{N}{4} and \Ion{N}{3} (middle), 
             \Ion{O}{4} and \Ion{O}{3} (bottom) lines
             of models calculated with 
             \Teffw{39, 42, 45} with the FUSE observation.
             We use models calculated with Kurucz' LIN lines in order to demonstrate the agreement with the
             local continuum.
            }
  \label{fig:Teff}
\end{figure}

The effective temperature \Teffw{42\pm 1} of \lb\ was determined by \citet{Rauch00} within small
error limits from the evaluation of ionization equilibria of 
\Ion{He}{1} / \Ion{He}{2}, \Ion{C}{3} / \Ion{C}{4}, \Ion{N}{3} -- \Ion{N}{5}, and \Ion{O}{4} / \Ion{O}{5}.
In the FUSE wavelength range, we also find metal lines of successive ionization stages, namely
\Ion{C}{3} / \Ion{C}{4}, 
\Ion{N}{3} / \Ion{N}{4}, 
\Ion{O}{3} / \Ion{O}{4},
\Ion{P}{4} / \Ion{P}{5}, and
\Ion{S}{4} / \Ion{S}{5}.
Due to the strong contamination by the ISM, many of these lines, e.g\@. the strong
\Ionw{C}{3}{977.03} and \Ionw{N}{3}{989.79} \sA{fig:spectrum} lines, cannot be used, but there are isolated lines
that are suitable for a analogous determination of \Teff\ 
(Figs\@. \ref{fig:phosphorus}, \ref{fig:sulfur}, \ref{fig:Teff}). 
Their ionization equilibria appear well-matched at \Teffw{42}, however, the strong contamination by ISM line absorption, 
the reddening, and the iron-group opacities \sK{sect:abundances} make it difficult to find the continuum flux 
level for a proper normalization. Thus, we estimate that our \Teff\ determination cannot be better than
$\pm 3\,\mathrm{kK}$. Consequently, we adopted \Teffw{42} for our analysis.

\subsection{Surface gravity}
\label{sect:logg}

To investigate the impact of the rotational velocity on the determination of the surface gravity of \lb,
we repeated the $\chi^2$ fit of \citet[][see his Fig\@. 4]{Rauch00} with the same synthetic and
observed fluxes, but we used $v_\mathrm{rot}=35\pm \mathrm{km/sec}$. We arrive now at a higher \loggw{5.30}
($\Delta \log g = 0.09$). These synthetic fluxes were calculated only from H+He models, so they suffer 
from the Balmer-line problem due to the neglecting metal opacities \citep{Werner96}. Consequently, 
we calculated a small grid of models that consider all elements from H -- Ni. Due to the relatively long 
calculation times of these much more detailed models, we had to restrict this grid to a fixed \Teffw{42}, 
\loggw{5.20\,-\,5.60} and $\Delta \log g = 0.05$. The $\chi^2$ fit yields \loggw{5.30} for
H\,$\alpha$ and H\,$\epsilon$ and \loggw{5.35} for H\,$\beta$, H\,$\gamma$, and H\,$\delta$. A comparison of
the observed hydrogen Balmer series with the observations \sA{fig:bergeron} shows clearly that a higher \logg\ does not
agree with the observations.

\begin{figure}[ht]
  \resizebox{\hsize}{!}{\includegraphics[]{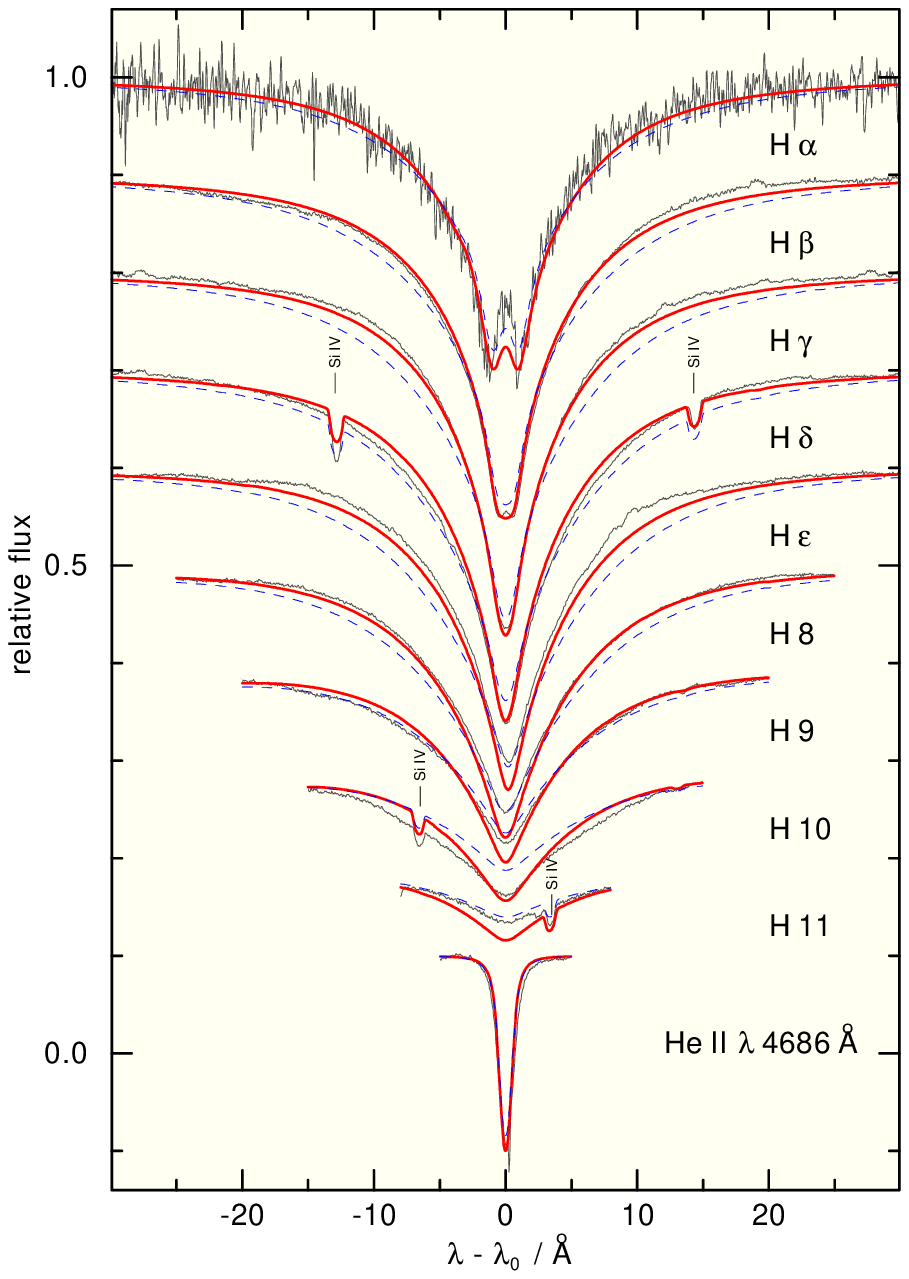}}
  \caption[]{Synthetic Balmer-line profiles calculated with \Teffw{42} and \loggw{5.3} (thick red line) and
             \loggw{5.6} (dashed, blue) compared with with our co-added (all 105 exposures) UVES observation.
             A surface gravity as high as \loggw{5.6} can definitely be excluded.
             The normalization of the observation to the continuum is not perfect for the
             broad Balmer lines.
             For the narrow \Ionw{He}{2}{4686} line, the \logg\ dependence is visible only in the innermost core.
            }
  \label{fig:bergeron}
\end{figure}

In the FUSE wavelength region, a higher surface gravity seems to improve the fit to the ``shoulders'' between the
higher hydrogen Lyman lines \sA{fig:logg_fuse}. However, the many uncertainties in the wavelengths range, e.g\@.
due to ISM absorption, unfortunately prevent any firm conclusion.

\begin{figure}[ht]
  \resizebox{\hsize}{!}{\includegraphics[]{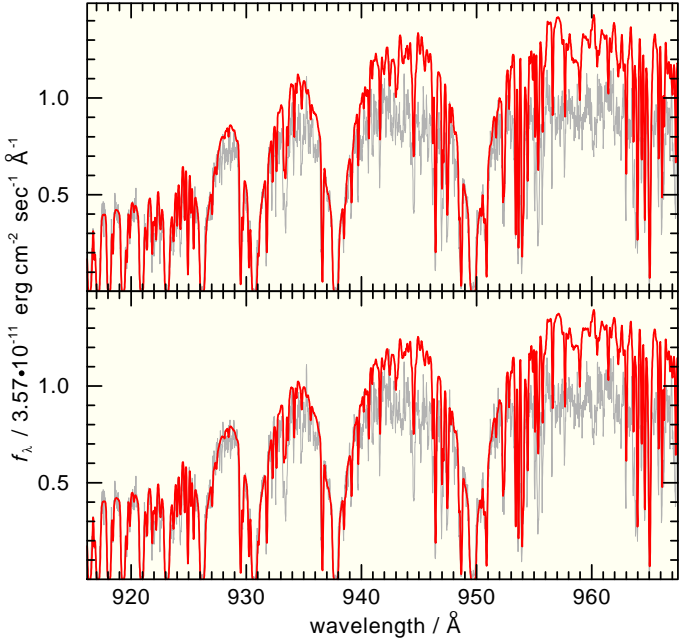}}
  \caption[]{Synthetic spectra calculated with \Teffw{42} and \loggw{5.3} (top panel) and
             \loggw{5.6} (bottom) compared with with our FUSE observation of \lb.
            }
  \label{fig:logg_fuse}
\end{figure}

\subsection{Abundances}
\label{sect:abundances}

The possibility of determining photospheric abundances from the FUSE observation is 
strongly limited by both the ISM line absorption \sA{fig:spectrum} and the limited number of
Kurucz' POS lines \sK{sect:analysis}. 
Only a few isolated photospheric lines are accessible in this wavelength range.
We adopted the previously determined abundances given by \citet{Rauch00} and the adjusted
Si and Ni abundances \sK{sect:NLTE}, because test calculations have shown that these give a 
good fit to the observation. 
The newly introduced elements phosphorus and sulfur enabled us to identify
lines of \Ion{P}{4} and \Ion{P}{5} \sA{fig:phosphorus} and \Ion{S}{4} and \Ion{S}{6} \sA{fig:sulfur}. 
The determined abundances are solar and 0.01 times solar, respectively.
Only weak lines of the iron-group element are located in the FUSE wavelength range.
Due to the stellar rotation \sK{sect:vrot}, we were not able to identify any of these lines.

\section{Results and conclusions}
\label{sect:results}

We performed an NLTE spectral analysis of FUSE observations of the post common-envelope binary \lb.
The short FUSE exposure times (200\,sec) allowed us to measure the rotational velocity 
$v_\mathrm{rot}=35 \pm 5\,\mathrm{km/sec}$ of the primary star of \lb. This is in good agreement with the
$v_\mathrm{rot}=34.7 - 38.6\,\mathrm{km/sec}$ given by \citet{Hilditch03}, and it confirms that the
rotation of \lb\ is bound.

A re-analysis of optical spectra \citep[cf\@.][]{Rauch00} has shown that \logg\ is about 0.1 higher than 
given by \citet[][\loggw{5.2}]{Rauch00}, who assumed bound rotation 
because the system has been classified to be a post common-envelope binary consisting of an sdOB and a 
main-sequence star 
\citep{deKoolRitter93} where the common-envelope phase is much longer than the synchronization time.
Since the primary's radius was assumed to be larger ($r_1=0.236\,\mathrm{R_\odot}$) due to the lower \logg, 
a higher $v_\mathrm{rot}=45\,\mathrm{km/sec}$ was adopted then for the analysis.
From $v_\mathrm{rot}=35\pm 5\,\mathrm{km/sec}$ and an orbital period of $P=22\,597.033\,\mathrm{sec}$ \citep{Kilkenny00}, 
we can calculate a stellar radius of $r_1=0.181\pm 0.025\,\mathrm{R_\odot}$. 

The spectral analysis is hampered by a strong ISM contamination. We used \owens\ to model the ISM line absorption 
qualitatively in order to identify and model stellar lines. Our NLTE models consider opacities of 18 species from H -- Ni.
It is obvious that the iron-group elements (here: Ca -- Ni) contribute to the opacity.
However, we were not able to identify any individual iron-group line in the FUSE observation.

We identified phosphorus and sulfur lines in the FUSE spectrum. 
For phosphorus we can determine a solar abundance. 
The sulfur abundance is surprisingly low, and we determined a 0.01 times solar abundance. 
The photospheric abundances of \lb\ are summarized in \ab{fig:abundances}.
The derived abundance pattern reflects the interplay of gravitational settling and
radiative levitation \citep[cf\@.][]{Rauch00}.

\begin{figure}[ht]
  \resizebox{\hsize}{!}{\includegraphics[]{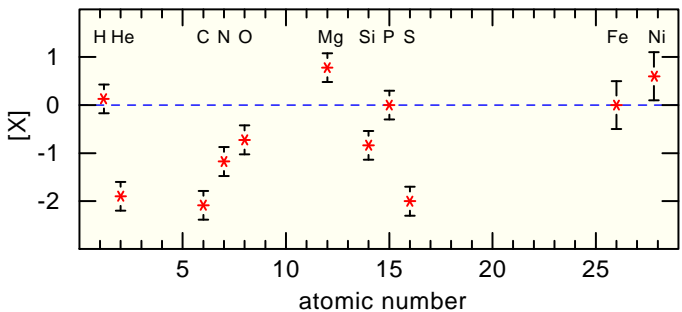}}
  \caption[]{Photospheric abundances of \lb\ determined by \citet{Rauch00} and in this work (Si, P, S, Ni). 
             [X] denotes $\log \mathrm{(mass~fraction / solar~mass~fraction)}$ of species X.}
  \label{fig:abundances}
\end{figure}

The ``gravity problem'' is still unsolved although the surface gravity is slightly higher: 
\loggw{5.3} rather than the 5.2 found by \citet{Rauch00}. 
If we take the radius of the primary as $r_1=0.181 R_\odot$ as noted above, and its mass $M_1=0.330M_\odot$ \citep{Rauch00}, 
then we obtain {\boldmath $\log g = \log \left(G M_1/R_1^2\right) = 5.44 \pm 0.13$\unboldmath} (where G is the gravitational constant).
Note that $\log g \approx 5.5$ is necessary for good agreement \citep[cf\@.][]{Rauch00}.

An additional uncertainty \citep[cf\@.][]{Rauch00} is the lack of appropriate evolutionary models
for post common-envelope binaries to compare so as to derive the primary's mass.

High-resolution and high S/N observations in the near-UV wavelength range is highly desirable when searching for the
weak lines of iron-group elements. \aba{fig:COS} demonstrates that lines of all elements from Ca -- Ni but
Sc and V should be detectable. The search for signatures of the ``heated-up'' secondary in IR observations in order to
further investigate the nature of this star also appears possible, and its contribution to the continuum at 
$\lambda = 12\,\mu$ is a few percent.

\begin{figure}[ht]
  \resizebox{\hsize}{!}{\includegraphics[]{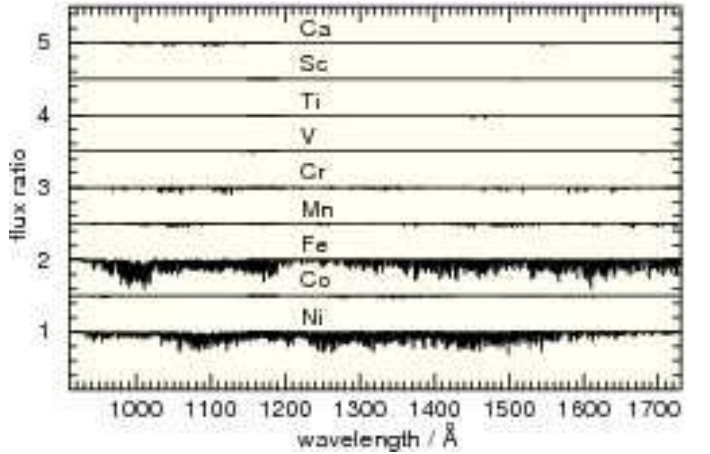}}
  \caption[]{Flux ratios (in the far UV) of synthetic spectra for \lb\ (calculated with abundances from
             \ab{fig:abundances} and Kurucz's LIN lines),
             which include only line opacities of one of the elements
             Ca -- Ni and a synthetic spectrum, which does not include any Ca -- Ni line.}
  \label{fig:COS}
\end{figure}

\begin{acknowledgements}
We like to thank our referee, Dave Kilkenny, for his useful comments that helped us to clarify and
improve this paper.
T.R\@. is supported by the \emph{German Astrophysical Virtual Observatory} (GAVO) project
of the German Federal Ministry of Education and Research (BMBF) under grant 05\,AC6VTB. 
J.W.K\@. is supported by the FUSE project, funded by NASA contract NAS5$-$32985.
This work was done using the profile-fitting procedure \owens\ developed by 
M\@. Lemoine and the FUSE French Team.
This work is based on INES\footnote{IUE Newly Extracted Spectra Archive Data Server} 
data from the IUE satellite.
Some of the data presented in this paper were obtained from the Multimission Archive at the 
Space Telescope Science Institute (MAST). STScI is operated by the Association of Universities 
for Research in Astronomy, Inc., under NASA contract NAS5-26555. Support for MAST for non-HST 
data is provided by the NASA Office of Space Science via grant NAG5-7584 and by other grants and contracts.
The UVES spectra used in this analysis were obtained as part of an ESO Service Mode run,
proposal 66.D$-$1800.
This research has made use of the SIMBAD database, operated at the CDS, Strasbourg, France.
\end{acknowledgements}

\bibliographystyle{aa}
\bibliography{0738.bbl}

\end{document}